\documentclass[aps,12pt,pra,showpacs,preprint,a4paper]{revtex4}

\usepackage{amsmath}
\usepackage{graphicx}
\usepackage{subfig}
\usepackage{array}
\usepackage{amsfonts}
\usepackage{epsf}
\usepackage{epsfig}
\usepackage{amssymb}
\usepackage{bbm}
\usepackage{delarray}

\begin{document}

\title{Treatment of $N$-dimensional Schr\"{o}dinger equation for anharmonic potential via Laplace transform}
\author{\small Tapas Das}
\email[E-mail: ]{tapasd20@gmail.com}\affiliation{Kodalia Prasanna Banga High
 School (H.S), South 24 Parganas, 700146, India}
\begin{abstract}
First time anharmonic potential $V(r)=ar^2+br-\frac{c}{r} \,,(a>0) $ is examined for $N$-dimensional Schr\"{o}dinger equation via Laplace transformation method. In transformed space, the behavior of the Laplace transform at the singular point of the differential equation is used to study the eigenfunctions and the energy eigenvalues.The results are easy to derive and identical with those obtained by other methods.

\end{abstract}
\maketitle
\newpage
\section{I\lowercase{ntroduction}}
Over the past few years, solving of non relativistic wave equation i.e Schr\"{o}dinger equation via Laplace transform has become one of the important research area specially for atomic and molecular physics. These type of studies have a long history which began from the first years of quantum mechanics by Schr\"{o}dinger when discussing the radial eigenfunction of the hydrogen atom [1]. More than forty years later Englefield approached the Schr\"{o}dinger equation with the Coulomb, oscillator, exponential
and Yamaguchi potentials [2]. Since then several research work has been carried out for other solvable potentials like pseudoharmonic [3-4], Morse like [5-6], Mie-type [7] and others [8-10]. The focus of these type of studies was to promote the method of Laplace transform over other parallel methods such that, series solution method [11], Fourier transfrom method [12], Nikiforov-Uvarov method [13], asymptotic iteration method [14], super-symmetric approach [15], shifted $\frac{1}{N}$ expansion method [16] and others[17-19].\\
Despite of the strong focus, we must say still there is a lack of study on Schr\"{o}dinger equation for anharmonic type potentials via Laplace transform. May be the main reason of this lack is due to the mathematical difficulties to obtain Laplace transformable differential equation from original wave equation. The term ``Laplace transformable" is meaningful here in the sense that inserting the chosen potential into Schr\"{o}dinger equation and imposing parametric restriction or by variable alteration we must get a ``modified differential equation" with coefficient say $r^j (j=0, 1)$ before triggering the Laplace transformation rule. Actually whenever $j>1$, eventually happens for anharmonic potentials, after Laplace transform of the modified differential equation we can not get first order differential equation in transformed space and hence privilege of mathematical easiness goes in vain. That is the reason why Laplace transform for any anharmonic potential is being avoided for a long time. Feeling the fact, in this paper first time a special anharmonic type potential  
\begin{eqnarray}
V(r)=ar^2+br-\frac{c}{r} \,,a>0 
\end{eqnarray}
is examined on $N$- dimensional Schr\"{o}dinger equation via Laplace transform. This potential is supposed to be responsible for the interaction between quark and antiquark and alternatively it is called \textit{Killinbeck} or \textit{Cornell plus harmonic potential}[20].The main motivation of this present work is to show that, Laplace transform may also provide mathematical comforts to tackle those modified differential equations which has variable coefficient for $j>1$. Here in this work we have encountered the situation for $j=2$ and also a fact of second order differential equation in transformed space instead of first order. Solution of such equation is easy in transformed space if the behavior of Laplace transform at the singular point is taken into account.
\section{{O\lowercase{verview Of}} L{\lowercase {aplace}} T{\lowercase {ransform}}}
The Laplace transform $\phi(s)$ or $\mathcal{L}$ of a function $f(t)$ is defined by [21]
\begin{eqnarray}
\phi(s)=\mathcal{L}\left\{{f(t)}\right\}=\int_{0}^{\infty}e^{-{st}}{f(t)}dt\,.
\end{eqnarray} 
If there is some constant $\sigma\in R$ such that ${\left|e^{-{\sigma}{t}}{f(t)}\right|\leq \bf{M}}$
for sufficiently large $t$, the integral in Eq.(2) will exist for  Re $s>\sigma$ . The Laplace transform may fail to exist because of a sufficiently strong singularity in the function $f(t)$ as $t\rightarrow 0$ . In particular
\begin{eqnarray}
\mathcal{L}\left[\frac{t^{\alpha}}{\Gamma(\alpha+1)}\right]=\frac{1}{s^{\alpha+1}}\,,{\alpha}>-1\,.
\end{eqnarray}
The Laplace transform has the derivative properties 
\begin{eqnarray}
\mathcal{L}\left\{f^{(n)}(t)\right\}=s^n\mathcal{L}\left\{f(t)\right\}-\sum_{k=0}^{n-1}s^{n-1-k}{f^{(k)}(0)}\,,
\end{eqnarray}
\begin{eqnarray}
\mathcal{L}\left\{t^{n}f(t)\right\}=(-1)^{n}\phi^{(n)}(s)\,,
\end{eqnarray}
where the superscript $(n)$ denotes the n-th derivative with respect to $t$ for $f^{(n)}{(t)}$, and with respect to $s$ 
for $\phi^{(n)}{(s)}$.
If $s_0$ is the singular point, the Laplace transformation behaves as for  $s\rightarrow s_0$
\begin{eqnarray}
\phi(s)=\frac{1}{(s-s_0)^{v}}\,,  
\end{eqnarray}
then for $t\rightarrow \infty$
\begin{eqnarray}
f(t)=\frac{1}{\Gamma(v)}t^{v-1}{e^{{s_0}{t}}}\,,
\end{eqnarray}
where $\Gamma(v)$ is the gamma function. On the other hand , if near origin $f(t)$ 
behaves like $t^\alpha$, with $\alpha>-1$, then $\phi(s)$ behaves near $s\rightarrow \infty$ as
\begin{eqnarray}
\phi(s)=\frac{\Gamma(\alpha+1)}{s^{\alpha+1}}\,.
\end{eqnarray}   
\section{B\lowercase{ound} s\lowercase{tate} s\lowercase{pectrum}}
In \textit{natural unit} $c=\hbar=1$,the $N$-dimensional time-independent Schr\"{o}dinger equation for a particle of mass $\mu$ with arbitrary angular quantum number 
$\ell$ is given by [22] 
\begin{eqnarray}
\Bigg[\nabla_N^{2}+2\mu\Big(E-V(r)\Big)\Bigg]\psi_{n\ell m}(r,\Omega_N)=0\,,
\end{eqnarray}
where $E$ and  $V(r)$ denote the energy eigenvalues and potential. $\Omega_N$ within the argument of $n$-th state eigenfunctions $\psi_{n\ell m}$ denotes angular variables
$\theta_1,\theta_2,\theta_3,.....,\theta_{N-2},\varphi$. The Laplacian operator in hyperspherical coordinates is written as 
\begin{eqnarray}
\nabla_N^{2}=\frac{1}{r^{N-1}}\frac{\partial}{\partial r}\left(r^{N-1}\frac{\partial}{\partial r}\right)-\frac{\Lambda_{N-1}^{2}}{r^2}\,,
\end{eqnarray}
where 
\begin{eqnarray}
\Lambda_{N-1}^{2}=-\Bigg[\sum_{k=1}^{N-2}\frac{1}{sin^2\theta_{k+1}sin^2\theta_{k+2}.....sin^2\theta_{N-2} sin^2\varphi}\times\left(\frac{1}{sin^{k-1}\theta_k}\frac{\partial}{\partial \theta_k}sin^{k-1}\theta_k\frac{\partial}{\partial \theta_k}\right)\nonumber\\+\frac{1}{sin^{N-2}\varphi}\frac{\partial}{\partial\varphi}sin^{N-2}\varphi\frac{\partial}{\partial\varphi}\Bigg]\,.
\end{eqnarray}
$\Lambda_{N-1}^{2}$ is known as the hyperangular momentum operator.\\ 
We chose the bound state eigenfunctions $\psi_{n\ell m}(r,\Omega_N)$ that are vanishing for $r\rightarrow0$ and $r\rightarrow\infty$. Applying the separation variable method by means of the solution
$\psi_{n\ell m}(r,\Omega_N)=R_{n\ell}(r)Y_\ell^{m}(\Omega_N)$,  Eq.(9) provides two separated equations
\begin{eqnarray}
\Lambda_{N-1}^{2}Y_\ell^{m}(\Omega_N)=\ell(\ell+N-2)Y_\ell^{m}(\Omega_N)\,,
\end{eqnarray} 
where $Y_\ell^{m}(\Omega_N)$ is known as the hyperspherical harmonics and the 
hyperradial or in short the ``radial'' equation 
\begin{eqnarray}
\left[\frac{d^2}{dr^2}+\frac{N-1}{r}\frac{d}{dr}-\frac{\ell(\ell+N-2)}{r^2}+2\mu[E-V(r)]\right]R_{n\ell}(r)=0\,,
\end{eqnarray}
where $\ell(\ell+N-2)|_{N>1}$ is the separation constant [23] with $\ell=0, 1,
2, \ldots$\\
Inserting the anharmonic potential into the Eq.(13) yields 
\begin{eqnarray}
\left[\frac{d^2}{dr^2}+\frac{N-1}{r}\frac{d}{dr}-\frac{\ell(\ell+N-2)}{r^2}+2\mu[E-(ar^2+br-\frac{c}{r})]\right]R(r)=0\,.
\end{eqnarray} 
In order to guess the eigenfunction at the asymptotic limit $r\rightarrow\infty$ we have 
\begin{eqnarray}
\frac{d^2{R}}{dr^2}-4{\alpha^2r^2}R=0\,,
\end{eqnarray}
where the fact $ar^2>>br$ for large $r$  has been assumed. The term $4\alpha^2$ is for mathematical convenience with   
\begin{equation}
\alpha=\sqrt{\frac{\mu{a}}{2}}\,.
\end{equation}
The solution of the Eq.(15) is easy. Remembering the fact $r^2>>1$, we can take the solution as
\begin{eqnarray}
R(r)\propto{e^{-\alpha r^2}}\,. 
\end{eqnarray} 
Now we are in a position to guess a complete solution of Eq.(14). The solution should be bounded in the origin. 
Let us try a solution of type
\begin{eqnarray}
R(r)=r^k e^{-\alpha r^2}{f(r)} \,,\;\;    k>0\,.
\end{eqnarray}
The term $r^k$ assures that, the solution at $r=0$ is bounded. The function $f(r)$ yet to be determined.
Inserting Eq.(18) into Eq.(14) we have
\begin{eqnarray}
rf^{''}(r)+(\eta_{kN}-4{{\alpha}r^2})f^{'}(r)
+\left\{\frac{Q(k,\ell,N)}{r}-2\mu{br^2}+2\mu{c}+r\epsilon_{kN}\right\}f(r)=0 \,,
\end{eqnarray} 
were the prime over $f(r)$denotes the derivative with respect to $r$. Also
\begin{eqnarray*}
\begin{aligned}
Q(k,\ell,N)&=k(k-1)+k(N-1)-\ell(\ell+N-2)\,,\\
\epsilon_{kN}&=2\mu E-2\alpha N-4\alpha k\,,\\
\eta_{kN}&=2k+N-1\,.
\end{aligned}
\end{eqnarray*}
To ease out the scenario of Laplace transform of the above differential equation we impose a parametric restriction 
\begin{eqnarray}
Q(k,\ell,N)=0\,.
\end{eqnarray}
This gives $k_+=\ell$ and $k_-=-(\ell+N-2)$ . The acceptable physical value of $k$ is  $k_+=\ell$.
Finally we have
\begin{eqnarray}
rf^{''}(r)+(\eta_{\ell N}-4{\alpha}r^2)f^{'}(r)
+(2\mu{c}-2\mu{br^2}+r\epsilon_{\ell N})f(r)=0 \,.
\end{eqnarray}
It is worth to mention here that Eq.(20) is not a mandatory condition of applying Laplace transformation on Eq.(21). It only helps to achieve the second order differential equation in transform space. If it was not imposed then the differential equation in transformed space would be of third order, and obviously we do not want that situation . Now we are in a position to get the Laplace transform of the above differential equation given by Eq.(21).Introducing the Laplace transform
$\phi(s)=\mathcal{L}\left\{{f(r)}\right\}$ and taking the boundary condition $f(0)=0$ , 
the derivative properties of Laplace transform Eq.(4)and Eq.(5) give
\begin{eqnarray}
(s+\beta)\frac{d^2\phi(s)}{ds^2}+\left(\frac{1}{4\alpha}s^2+\lambda\right)\frac{d\phi(s)}{ds}+\left(\gamma{s}-\frac{\mu{c}}{2\alpha}\right)\phi(s)=0\,,
\end{eqnarray}
where the following abbreviations are used
\begin{eqnarray}
\beta=\frac{\mu{b}}{2\alpha} \,\,; \lambda=\frac{\mu{E}}{2\alpha}-(\ell+\frac{N}{2}-2) \,\,;  \gamma=\frac{3-2\ell-N}{4\alpha}\,. 
\end{eqnarray}
The singular point of transformed differential equation is $s=-\beta$ \\
Using the fact of Eq.(6) the solution of Eq.(22)can be taken
\begin{eqnarray}
\phi(s)=\frac{C_{n\ell N}}{(s+\beta)^{n+1}} \,, n=0,1,2,3, \ldots
\end{eqnarray}
Inserting the Eq.(24) in Eq.(22) we get the following three identity relations,
\begin{eqnarray}
\gamma=\frac{n+1}{4\alpha}\,,
\end{eqnarray}
\begin{eqnarray} 
\gamma\beta=\frac{\mu{c}}{2\alpha}\,,
\end{eqnarray}
\begin{eqnarray}
(n+1)(n+2)-(n+1)\lambda-\frac{\mu{c}}{2\alpha}{\beta}=0\,.
\end{eqnarray}
Using the set of Eq.(25-27) and Eq.(23), the energy eigenvalues becomes
\begin{eqnarray}
E_{n\ell N}=\sqrt{\frac{a}{2\mu}}(2n+2\ell+N)-\frac{b^2}{4a}\,.
\end{eqnarray} 
This result is exactly the same that obtained in [24-27] before. 
The undetermined function $f(r)$ can be found from the inverse Laplace transform such that 
$f(r)=\mathcal{L}^{-1}\{\phi(s)\}$.\\ 
Using the Eq.(7), we get
\begin{eqnarray}
f(r)=\frac{C_{n\ell N}}{n!}r^{n}e^{-\beta{r}}\,.
\end{eqnarray}
Finally using Eq.(16), Eq.(18) and Eq.(23) it is easy to determine the corresponding eigenfunctions of the system as 
\begin{eqnarray}
R_{n\ell N}(r)=\frac{C_{n\ell N}}{n!}r^{\ell+n}\exp{(-\sqrt{\frac{\mu{a}}{2}}r^2-b\sqrt\frac{\mu}{2a}r)}\,.
\end{eqnarray}
The normalization constant $C_{n\ell N}$ can be obtained from the condition 
\begin{eqnarray}
\int_{0}^{\infty}[R_{n\ell N}(r)]^2r^{N-1}dr=1\,.
\end{eqnarray} 
A fare approximation $r+\frac{\beta}{2\alpha}\approx r$ will make the integration much more easy to evaluate. The use of integral formula 
\begin{eqnarray*}
\int_{0}^{\infty}x^p e^{-Ax^q}dx=\frac{1}{q}\frac{\Gamma(\frac{p+1}{q})}{A^{\frac{p+1}{q}}}\,, \;\;    p,q>0\,, 
\end{eqnarray*}
gives 
\begin{eqnarray}
C_{n\ell N}=n!\left\{\frac{2(2\alpha)^{\ell+n+\frac{N}{2}}}{\Gamma(\ell+n+\frac{N}{2})}e^{-\frac{\beta^2}{2\alpha}}\right\}^{\frac{1}{2}}\,.
\end{eqnarray}
In case of three dimensional isotropic harmonic oscillator $V(r)=\frac{1}{2}\mu \omega r^2=ar^2$ and $b=c=0$. Within the frame work of natural unit, Eq.(28) gives the energy eigenvalue $E=\omega (n^{'}+\frac{3}{2})$, which is a very common result with the definition of quantum number $n^{'}=n+\ell$.  
\section {c\lowercase{onclusion}}
In this paper $N$-dimensional Schr\"{o}dinger equation for anharmonic potential has been solved via Laplace transformation. The differential equation in transformed space is second order instead of the first order. It has been shown that this is not the problem at all for deriving energy eigenvalues as well as eigenfunctions. Moreover we can say after achieving the second order differential equation the method of deriving energy eigenvalues and eigenfunctions are easier than the usual guide line of conventional method of Laplace transformation i.e trying a first order differential equation in transformed space and then imposing a quantization condition on the solution of that.\\
The potential $V(r)=ar^2+br-\frac{c}{r} \,,(a>0)$ has been solved by other methods as cited earlier. Mass spectra of quarkonium system  $(\Upsilon(b\overline{b}),\psi(c\overline{c}))$ are well studied using equation Eq.(28). The only important thing in this paper is to show an example that Laplace transformation is equally a powerful method to tackle anharmonic potential. Furthermore, from present calculation using the approximation $r+\frac{\beta}{2\alpha}\approx r$,the approximate \textit{rms} radius of the bound state of quarkonium comes out as  
$r_{rms}=\sqrt{<r^2>}=\sqrt{\frac{\ell+n+\frac{3}{2}}{2\alpha}}\times0.1973 fm,$ 
where $<r^2>=\int_{0}^{\infty}r^2[R_{n\ell 3}(r)]^2r^2dr.$ Careful calculation shows that the \textit{rms} radius of $b\overline{b}$ for the state $1S(\Upsilon)$ is $0.2672fm$ and the same for $c\overline{c}$ for the state $1S(J/\psi)$ is $0.4839 fm$. These results are quite a good matched with reported values. All other \textit{rms} radii for higher states are somehow differ from the actual value of the present day experimental results. Despite of non relativistic model, this achievement is also remarkable thing to note down.\\
There are several anharmonic potentials like \textit{Singular integer power potential} $V(r)=\sum_{i=0}^p \frac{a_i}{r^i}$, \textit{Sextic potential} $V(r)=ar^6+br^4+cr^2$, \textit{Non-polynomial potential} $V(r)=r^2+\frac{\lambda r^2}{1+gr^2}$ and others. These potentials are still untouched by the use of Laplace transformation. Though there are other mathematical questions about the applicability of Laplace transformation for every potential but the present paper will help little bit to overcome the barrier and strict guide line of obtaining the first order differential equation in transformed space. 
\section*{R\lowercase{eferences}}    

\end{document}